\documentclass[12pt]{article}
\usepackage{graphicx}

\begin{document}

\begin{center}
\Large {\bf Integration  of Constraint Equations in Problems of a Disc and a Ball Rolling on
a Horizontal Plane}
\end{center}

\bigskip
\bigskip

\begin{center}
Eugeny A. Mityushov
\end{center}

\bigskip
\bigskip

\begin{center}
{\it Ural Federal University\\
     Department of Theoretical Mechanics\\
     Prospect Mira 19\\
     620 002 Ekaterinburg\\
     Russian Federation\\}
 \medskip
\end{center}

\begin{center}
e-mail:~ mityushov-e@mail.ru
\end{center}

\bigskip
\bigskip

\begin{abstract}
The problem of a  disc and a ball rolling on a horizontal plane without slipping is
considered. Differential constrained equations are shown to be integrated when the
trajectory of the point of contact is taken in a form of the natural equation, i.e. when
the dependence of the curvature of the trajectory is explicitly expressed in terms of
the distance passed by the point.
\end{abstract}

\newpage

\section{The problem of a rolling disc}
\setcounter{equation}0

The rolling motion of a disc and a ball on a static plane was described many times, see
e.g.  \cite{K6,K7}, and generalized in later publications \cite{K12}-\cite{K19}. These
are the classical examples of motion of mechanical systems with non-holonomic
constraints.

\begin{figure}[h]
\begin{center}
\includegraphics[height=5cm,width=0.5\textwidth]{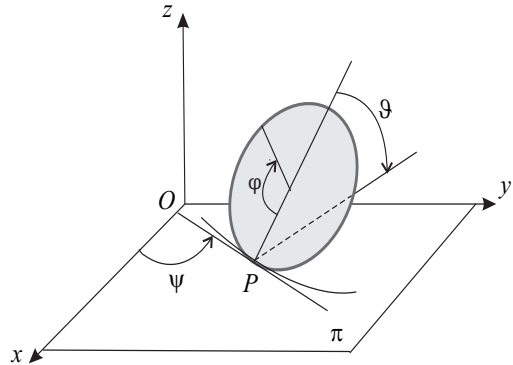}
\caption{\small{Generalized coordinates of a rolling disk.}}
\label{f1}
\end{center}
\end{figure}

Let a disc with radius $R$ be tangent to a plane $\pi$ with the system of coordinates
$Oxy$, let $P$ be the point of contact between the disc and the plane. The position of a
disk is determined by five independent coordinates. For example, it can be fixed by
coordinates  $x_{P}$ and $y_{P}$, by the angle of rotation $\varphi$, by the angle of
precession $\psi$, and by the angle of nutation $\vartheta$, see Figure \ref{f1}.

Evolution of the coordinates always satisfies the two non-holonomic constraints which
can be written in the following form:
\begin{equation}\label{4}
dx_{P} = R\cos\psi d\varphi~~,~~dy_{P}=R\sin\psi d\varphi~~.
\end{equation}

The problem above is, therefore, to integrate the given dynamical system under applied
arbitrary external forces. The problem is interesting because the methods of the
classical mechanics it requires to use are  rather involved. The solution can be
essentially simplified if the rolling trajectory is known. Such a formulation is
possible, for example, under modeling of the rolling with the help of a computer
animation.

In fact, the kinematics of the rolling motion of the disk is determined by the three
functions
\begin{equation}\label{1}
\varphi=\varphi(t)~~,~~\vartheta=\vartheta(t)~~,~~\psi=\psi(t)~~.
\end{equation}
Under the rolling motion without slipping the arc coordinate $s$ of the point of contact
of the disc and the plane is related  to the angle of rotation $\varphi$ as
\begin{equation}\label{5}
\varphi=s/R~~.
\end{equation}
Therefore,
\begin{equation}\label{6}
{\dot{\psi} \over \dot{\varphi}}={Rk(s)}~~.
\end{equation}
Here $k(s)=\frac{d\psi}{ds}$ is a curvature of the point of contact trajectory for the
disc.

It is a well-known that given any function  $k=k(s)$ one can find a curve
$\vec{r}=\vec{r}\,(s)$ with the curvature equal $k(s)$. The curve is unique up to a
congruence. Equation $k=k(s)$ is known as a natural equation of the curve. The
parametric equations of the trajectory of the point of contact,
\begin{equation}\label{2}
x_{P} = \int_0^s\cos(\int_0^\tau k(s)ds)d\tau ,~~~y_{P} = \int_0^s\sin(\int_0^\tau
k(s)ds)d\tau~~,
\end{equation}
see \cite{K21}, allow one to find the location of the disk on the plane at any moment of
the motion.

Equations (\ref{1}) enable one to describe kinematics of the disk with equations of the
point of contact in form (\ref{2}). The found solution satisfies non-holonomic
constraint equations (\ref{4}).

Let us give an example of how rolling of the disc can be described by three equations of
motion (\ref{1}) by using the natural equation for the trajectory of the point of
contact with a  horizontal plane.

{\bf Example 1:} Let rolling of the disk be determined by following equations:
\begin{equation}\label{3}
\varphi=\omega t,~~~\psi=\frac{\displaystyle \varepsilon t^2}{2},~~~\vartheta=f(t)~~.
\end{equation}
The curvature of the trajectory of the point of contact is
\begin{equation}\label{7}
k(s)={\dot{\psi} \over \dot{s}}={\varepsilon t \over  R \omega}={\varepsilon s \over  (R
\omega)^{2}}~~,
\end{equation}
where $\varepsilon$ and $\omega$ are some parameters, $f(t)$ is a function. This
trajectory is the clothoid whose asymptotic point has coordinates $x=y=(R\omega\sqrt{\pi
/\varepsilon})/2$.

The equations of motion of the point of contact in this case are
\begin{equation}\label{8}
x_{P} = \int_0^{s}\cos\frac{ \varepsilon s^2}{ 2(R\omega)^2}ds~~,~~
y_{P} = \int_0^{s}\sin\frac{\varepsilon s^2}{ 2(R\omega)^2}ds~~;
\end{equation}
or
\begin{equation}\label{9}
x_{P} = R\omega\int_0^{ t} \cos\frac{\varepsilon t^2}{ 2}dt~~,~~
y_{P} = R\omega\int_0^{ t} \sin\frac{\varepsilon t^2}{ 2}dt~~.
\end{equation}

A direct substitution of these functions into the equations (\ref{1}) shows that
obtained solution satisfies the constraints.

\begin{figure}[h]
\begin{center}
\includegraphics[width=0.9\textwidth]{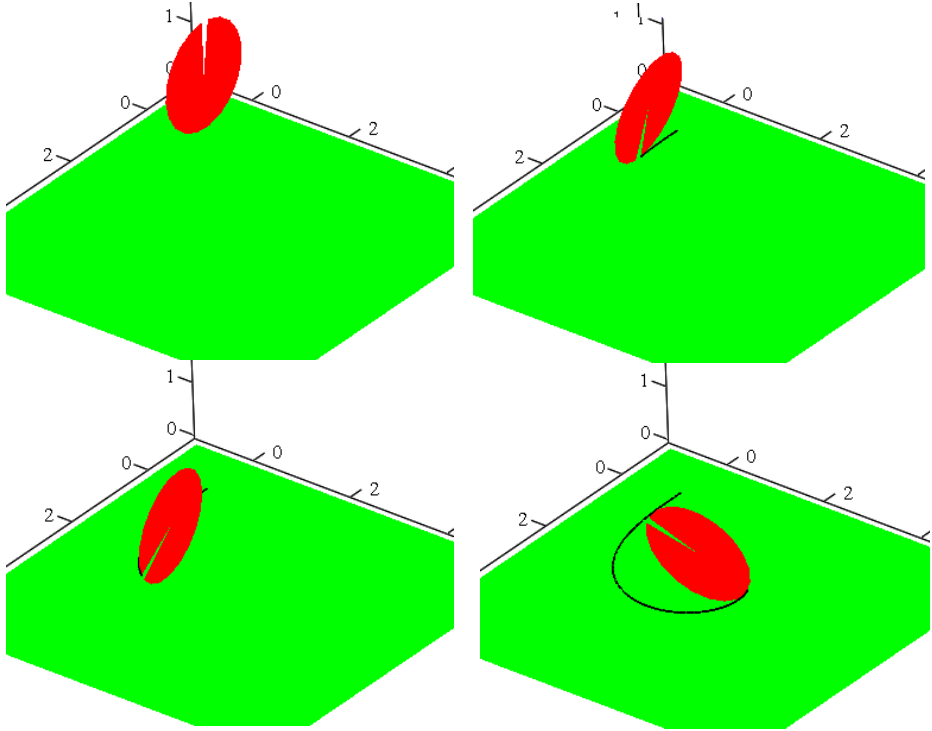}
\caption{\small{Phases of motion of the disk along a clothoid
$(\omega=\pi, \varepsilon=\frac{\pi}{16}, t^{*}=16)$}} \label{f2}
\end{center}
\end{figure}

The phases of motion are shown on Fig. \ref{f2}. They are obtained by using equation
(\ref{3}) with $f(t)=\frac{\displaystyle\pi}{\displaystyle t^{*2}}(t-t^*)^2.$

\section{The problem of a rolling ball}

\begin{figure}[h]
\begin{center}
\includegraphics[width=0.6\textwidth]{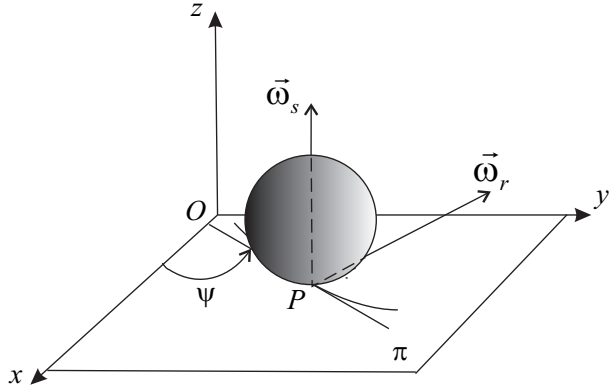}
\caption{\small{The rolling and spinning angular velocity
vectors of the ball.}} \label{f3}
\end{center}
\end{figure}

Let us now discuss a similar problem for a ball. We assume that the ball rolls on a
plane and spins simultaneously. Like in case of the disc, the position of a ball (see
Fig. \ref{f3}) can be determined by three functions. To introduce these functions we
decompose the vector of angular velocity of the ball  $\vec{\Omega}$ into two parts, as
shown on Fig. \ref{f3},
\begin{equation}\label{10}
\vec{\Omega}=\vec{\omega}_{s}+\vec{\omega}_{r}~~.
\end{equation}
The vector of angular velocity related to the spinning, $\vec{\omega}_{s}$,
is a orthogonal to the plane. The vector of the angular velocity associated to the
rolling, $\vec{\omega}_{r}$, is parallel to the plane.

Evolution of the  velocity vectors at the center of the ball is determined by the same
angle $\psi$. By taking this into account one can find the complete set of functions
which describe the motion of the ball:
\begin{equation}\label{11}
\varphi=\int_0^t \omega_{r}(t)dt~~,~~~ \chi=\int_0^t \omega_{s}(t)dt~~,~~~
\psi=\psi(t)~~.
\end{equation}
Angles $\varphi$, $\chi$, $\psi$ are quasi-coordinates, they do not enable one
to establish position of the ball
in different moments of time. The position can be fixed by integrating constraint equations which are
identical to (\ref{4}). A solution to these equations formally coincides with equalities (\ref{2}).

It allows one to find the position of the ball at
any moment of motion. For the motion without spinning the position of the
ball is determined by only two functions $\varphi(t)$ and $\psi(t)$. Consider again an
example.

{\bf Example 2:} Let the ball of the radius  $R$ roll according to the following
equations:
\begin{equation}\label{12}
\varphi=\omega_{1}t,\qquad \psi=\omega_{2}t~~,
\end{equation}
where $\omega_{1}$ and $\omega_{2}$ are the corresponding angular velocities which are
assumed to be some constants. The curvature of the trajectory of the point of contact
with the plane is a constant,
\begin{equation}\label{13}
k=\frac{\dot{\psi}}{\dot{s}}=\frac{\omega_{2}}{R\omega_{1}}.
\end{equation}
This means that the point of contact of the ball with the plane and its center moves
along a circle with the radius $R\omega_{1}/\omega_{2}$. The equations of the circle
are
\begin{equation}\label{14}
x_{P}=\frac{R\omega_{1}}{\omega_{2}}\sin\omega_{2}t,\qquad
y_{P}=\frac{R\omega_{1}}{\omega_{2}}(1-\cos\omega_{2}t).
\end{equation}
The motion of other points of the ball can be determined by the Euler formula which
defines a velocity $\vec{v}$ for a point of a ball
\begin{equation}\label{15}
{\vec{v}}=[\vec{\omega}_{r},\vec{r}- \vec{r}_{p}]~~,
\end{equation}
where $\vec{\omega}_{r}=\{-\sin\omega_2 t,\cos\omega_2 t,0\}$.

By using (\ref{14}), (\ref{15}) one finds
\begin{equation}\label{16}
\frac{d{\bf r}}{dt}=\omega_{1} \left( \begin{array}{c} z\cos\omega_{2}t\\[4pt]
z\sin\omega_{2}t\\[4pt]-x \cos\omega_{2}t-y
\sin\omega_{2}t+\frac{R\omega_{1}}{\omega_{2}}\sin\omega_{2}t\\
\end{array}
\right).
\end{equation}

\begin{figure}[h]
\begin{center}
\includegraphics[width=0.9\textwidth]{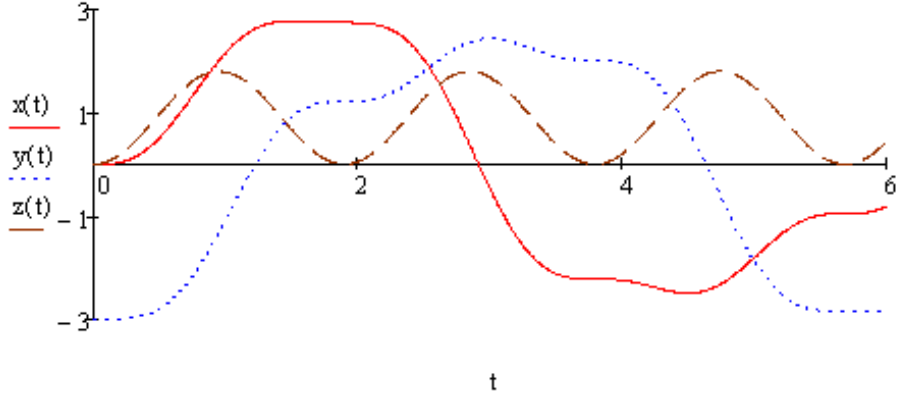}
\caption{\small{The behavior of the "south" pole of the ball under rolling along the
circle $(\omega_{1}=\pi~,~ \omega_{2}=\frac{\pi}{3}~,~ R=1)$.}} \label{f4}
\end{center}
\end{figure}

In coordinates with the origin at the center of the circle trajectory (of the point of
contact) this equation takes the form
\begin{equation}\label{17}
\frac{d{\bf r}}{dt}=\omega_{1} \left( \begin{array}{c} z\cos\omega_{2}t\\[4pt]
z\sin\omega_{2}t\\[4pt]-x \cos\omega_{2}t-y \sin\omega_{2}t\\
\end{array}
\right).
\end{equation}
One immediately checks that one of the integrals of this equation is expressed as
\begin{equation}\label{18}
x\frac{\omega_{1}}{\omega_{2}} \sin\omega_{2}t-y\frac{\omega_{1}}{\omega_{2}}
\cos\omega_{2}t+z=C~~.
\end{equation}
Fig. \ref{f4} shows the behavior of $x$, $y$, $z$ coordinates of a point of the ball being
initially a "south" pole of the ball.

\section{Conclusions}

Our analysis shows that in problems of a disk and a ball rolling on a plane it is a
sofficient to choose three functions to determine the laws of motion (\ref{2}), (\ref{11}).
(The choice of other parameters is also possible.) For the disc these functions are
$x_{P}=x_{P}(t)$, $y_{P}=y_{P}(t)$, and $\vartheta= \vartheta(t)$. For the ball they are
$x_{P}=x_{P}(t)$, $y_{P}=y_{P}(t)$, and $\chi=\chi(t)$. Other coordinates which fix
orientations of the given bodies are determined by coordinates of the point of contact
\begin{equation}\label{19}
\psi=\arccos\frac{\dot{x}_P}{\sqrt{\dot{x}_{P}^{2}+\dot{y}_{P}^{2}}}~~~,~~~\varphi=\frac{1}{R}\int_0^{t}
\sqrt{\dot{x}_{P}^{2}+\dot{y}_{P}^{2}}dt~~.
\end{equation}

This form of kinematic equations of motion is essentially convenient when the
trajectories of the rolling are given.

The following statement is an important consequence of the considered kinematic
description: a vertical rolling of a disc or a (non spinning) ball on a plane is
completely determined by (or equivalent to) the motion of a point (of the same mass) on
a plane under the force equal to the main vector of external forces applied to rolling
bodies.

The validity of this statement is based on Eq. (\ref{19}) and on the equivalence of
equations of motion of a point on a plane to equations of motion of the center of mass
of a disc or a ball under the action of the same system of forces.


\newpage








\begin{thebibliography}{}



\bibitem{K6}
 Appell P.{\it Theoretical mechanics 2}, Paris: Gauthier-Villars, 1953, 487p.

\bibitem{K7} Pars L.A. {\it A Treatise on Analytical Dinamics}, London: Heinemann, 1964,
    635p.


 \bibitem{K12}
Cushman, R. Routh's sphere. Rep. Math. Phys., 1998, 42, 47-70.

\bibitem{K13} O'Reilly, O. M. {\it The Dynamics of Rolling Disks and Sliding Disks},
    Nonlinear Dynamics, 1996, 10, 287-305.

\bibitem{K14} Hermans, J.  {\it A symmetric sphere rolling on a surface}, Nonlinearity,
    1995, 8, 493-515.

\bibitem{K15} Zenkov, D. V. {\it The geometry of the Routh problem}, J. Nonlinear Sci.,
    1995, 5, 503-519.

\bibitem{K16} Schneider, D. {\it Non-holonomic Euler-Poincare equations and stability in
    Chaplygin's sphere}, Dynamical Systems, 2002, 17, 87-130.


\bibitem{K19} Borisov A. V., Mamaev I.S. {\it Conservation Laws, Hierarchy of Dynamics
    and Explicit Integration of Nonholonomic Systems}, Nonlinear Dynamics , 2008,  v.4,
    3, 223-280.


\bibitem{K21} Struik, D.J. {\it Lectures on Classical Differential Geometry}, New York:
    Dover, 1988,~225p.


\end{thebibliography}
\end{document}